# Local analogues of high-redshift star-forming galaxies: integral field spectroscopy of green peas

E. K. Lofthouse[1]*, R. C. W. Houghton[2] and S. Kaviraj[1]

[1] *Centre for Astrophysics Research, School of Physics, Astronomy and Mathematics, University of Hertfordshire, College Lane, Hatfield, AL10 9AB, UK*
[2] *Dept. of Physics, University of Oxford, Denys Wilkinson Building, Keble Road, Oxford OX1 3RH*

26 June 2017

**ABSTRACT**
We use integral field spectroscopy, from the SWIFT and Palm3K instruments, to perform a spatially-resolved spectroscopic analysis of four nearby highly star-forming 'green pea' (GP) galaxies, that are likely analogues of high-redshift star-forming systems. By studying emission-line maps in H$\alpha$, [NII]$\lambda\lambda$6548,6584 and [SII]$\lambda\lambda$6716,6731, we explore the kinematic morphology of these systems and constrain properties such as gas-phase metallicities, electron densities and gas-ionization mechanisms. Two of our GPs are rotationally-supported while the others are dispersion-dominated systems. The rotationally-supported galaxies both show evidence for recent or ongoing mergers. However, given that these systems have intact disks, these interactions are likely to have low mass ratios (i.e. minor mergers), suggesting that the minor-merger process may be partly responsible for the high SFRs seen in these GPs. Nevertheless, the fact that the other two GPs appear morphologically undisturbed suggests that mergers (including minor mergers) are not *necessary* for driving the high star formation rates in such galaxies. We show that the GPs are metal-poor systems (25-40 per cent of solar) and that the gas ionization is not driven by AGN in any of our systems, indicating that AGN activity is not co-eval with star formation in these starbursting galaxies.

## 1 INTRODUCTION

Understanding the processes that drive galaxy evolution, particularly at high redshift, is a key topic in observational astrophysics and remains a subject of much debate. The star formation rate (SFR) density has changed significantly over cosmic time. While the average SFR in the nearby Universe is relatively low, it peaked around $z = 2 - 3$, when it was more than an order of magnitude greater than it is at low redshift (e.g. Madau, Pozzetti & Dickinson 1998; Hopkins & Beacom 2006). Much of the stellar mass seen in today's galaxies was therefore assembled at this epoch. Thus, to understand galaxy evolution over cosmic time, we need to understand the processes that were driving star formation, morphological transformation and black-hole growth around the epoch of peak cosmic star formation. However, it is difficult to perform detailed spatially-resolved studies of galaxies at this epoch due to their high redshift. A useful alternative is to study local analogues of the high-redshift galaxy population i.e. galaxies in the nearby Universe that have similar properties to that of typical galaxies found at earlier cosmic times (e.g. Overzier et al. 2013; Bian et al. 2016; Cowie, Barger & Songaila 2016;

Faisst et al. 2016; Greis et al. 2016). Detailed studies of such systems can offer valuable insights into the processes that drive galaxy evolution in the early Universe.

One class of objects that are plausible analogues of high redshift galaxies are the population of so-called 'green peas' (GPs), first detected as part of the Galaxy Zoo project (Cardamone et al. 2009) and so named because they have a compact, greenish appearance in the SDSS imaging. Although these galaxies have low stellar masses ($10^{8.5}$ - $10^{10} M_\odot$) they exhibit very high star formation rates (SFRs) of up to $\sim 10 M_\odot$/yr (Cardamone et al. 2009). If this rate of star formation is maintained, these galaxies are able to double their stellar mass within only $\sim$100 Myrs. The high SFRs in these galaxies produce strong emission lines with equivalent widths of $\sim$1000Å. In particular, the bright [OIII]$\lambda$5007 line falls within the $r$-band, which is coloured green in the SDSS *gri* composite images, giving these galaxies their greenish colours. They also happen to lie at $z \sim 0.2 - 0.3$ where they are unresolved in the SDSS imaging, which gives them their compact appearance. Fig. 1 shows images of the GPs studied in this paper.

Given their high SFRs and low masses, GPs fall at the top of the local sSFR-mass sequence making them some of the most actively star-forming galaxies in the local Universe. Although these sSFRs are typically extreme at low redshift, the star-forming properties of the GPs are comparable to those seen in galaxies in the early Universe, where star formation is more vigorous and such sSFRs are routinely observed (e.g. Daddi et al. 2010; van der Wel et al. 2011; Rodighiero et al. 2011; Kaviraj et al. 2013; Maseda et al. 2014; De Barros, Reddy & Shivaei 2016; Amorín et al. 2017). They also have compact sizes, ionization parameters and electron densities similar to these high-redshift galaxies (Bian et al. 2016), exhibit low dust extinction (Cardamone et al. 2009) and relatively low metallicities (Izotov, Guseva & Thuan 2011; Amorín et al. 2012a), with some showing spectral signatures of Wolf-Rayet stars (Amorín et al. 2012a). Taken together, these properties make the GPs good local analogues for galaxies in the early Universe, offering us a window into understanding the physics of star formation at early epochs.

* E-mail: e.k.lofthouse@gmail.com

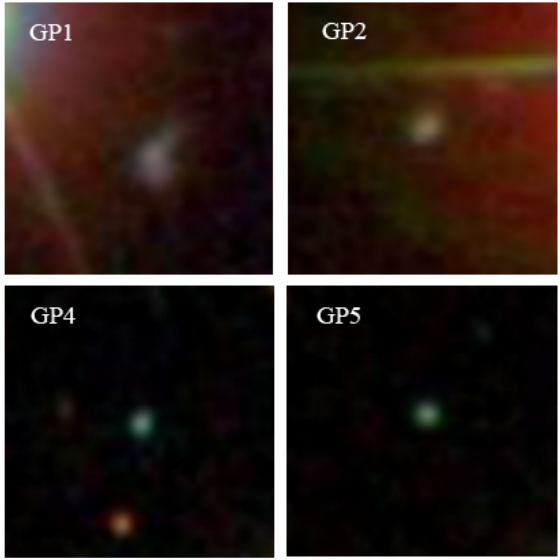

**Figure 1.** SDSS images of the four GPs (GP1, GP2, GP4 and GP5) targeted with SWIFT+Palm3K. The images are 15″ x 15″ in size.

It is still unknown what processes might influence the strong SFRs observed in the GPs. Archival HST images of some GPs indicate disturbed morphologies, suggesting that recent mergers may be responsible for inducing the high SFRs observed in at least some of these systems (Cardamone et al. 2009). In this study, we use integral field spectroscopy (IFS) to perform a resolved spectroscopic analysis of the internal structure of four of these systems, which provides a more detailed picture of their evolution than is possible using only globally averaged properties. Such an analysis offers insights into the kinematic morphology of these systems and enables us to put constraints on properties such as gas-phase metallicities and ionization mechanisms in a spatially-resolved way. In turn, this allows us to explore the physical conditions within these galaxies, the processes that drive their high SFRs and the potential role of AGN in regulating this star formation. This paper is structured as follows. In Section 2, we describe the selection of our target GPs and the SWIFT+PALM3K Integral Field Unit (IFU) observations. In Section 3, we present maps of the kinematics and emission line properties of our four GPs and in Section 4 we discuss our results and explore the nature of each galaxy. We summarize our findings in Section 5. Throughout the paper, we use the following cosmological parameters: $\Omega_M = 0.241$, $\Omega_\Lambda = 0.759$ and h = 0.732.

## 2    DATA

The targets in this study were selected from the sample of Cardamone et al. (2009) to have bright (V<15 mag) and nearby (r<25″) single conjugate adaptive optics reference stars (SCAORSs). In total, five targets were found that could be observed from Palomar and are listed in Table 1. However, due to time and weather constraints, only four were observed with sufficient signal to be useful: GP1, GP2, GP4 and GP5 (see Fig 1). Fig. 2 shows a comparison between the GPs and the sSFR-mass relations at various redshifts. The overall population of green peas is consistent with the sSFR and stellar of galaxies at high redshifts. The three GPs studied in this work are consistent with lower redshift sSFR-mass sequences than the average GP but there is significant scatter in these observational sequences. However, the GPs do fall on mass-metallicity relations expected for high redshift galaxies as shown in Fig. 3 and discussed in more detail in Section 3.3.

The Oxford Short Wavelength Integral Field SpecTrograph (SWIFT, Thatte et al. 2006) instrument is a visible light IFU at the 200 inch (5.1m) Hale Telescope at Palomar, offering spectroscopy at $0.63\mu$m-$1.04\mu$m. Combined with the PALM3K adaptive optics system (Dekany et al. 2013), we are able to study the resolved properties of our sample of GPs. SWIFT has a choice of three plate scales and fields of view (FOV). Although the data were observed with adaptive optics assistance, the faint magnitudes and large separations of the SCAORSs mean that the AO correction will not be diffraction limited. PALM3K is an extreme-AO system, designed to reach the diffraction limit of the telescope with some of the brightest stars in the sky (V$\gtrsim$7 mag). PALM3K can operate with SCAORSs as faint as V=17 mag, but with significantly reduced performance. With our SCAORSs, (12$\lesssim$V$\lesssim$15) we expect to improve on the seeing to sub arcsecond levels, but do not expect to reach resolutions near the diffraction limit. For this reason, and to maximise sensitivity, we chose to observe in the coarsest 235mas scale, giving a FOV of 10 arcsec × 20 arcsec.

The SWIFT data were reduced using the SWIFT data reduction pipeline (v2.5.0), written for IRAF. The pipeline includes standard CCD data reduction steps such as bias subtraction (using a fit to the overscan region), wavelength calibration (using Neon and Argon lamp exposures) and flat fielding (using halogen lamp exposures, with columns normalised by a Chebyshev polynomial) as well as IFS-specific stages such as cube reconstruction (using a vertical line mask illuminated by the halogen lamp) and illumination correction (using the halogen lamp exposures divided by the shape of the halogen spectrum). The final wavelength calibration is accurate to better than 0.1Å. Cosmic rays were removed using the LACOSMIC routine (van Dokkum 2001). We correct for flexure along the spectral axis using the night sky emission lines. Targets were observed using the standard near-infrared ABBA technique, and adjacent frames were subtracted and inverted to make A-B and B-A frames; these were then shifted and co-added using offsets derived from 2D Gaussian fits to the Hα flux maps of the target. We estimate the spatial resolution of our SWIFT+PALM3K observations directly from the data themselves. We assume that the bright Hα emission in the centre of each galaxy is unresolved and its FWHM then defines the FWHM of our AO-assisted PSF. For GP1 & GP2, we measure a FWHM of 0.5 arcsec, while for GP4 & GP5 we measure a FWHM of 0.8 arcsec.

Prior to the emission-line analysis, the data cubes are smoothed in QFitsView using the boxcar median method along X and Y axes with a smoothing width of 3 spaxels. These smoothed maps are used for the analysis of the galaxy kinematics, electron densities and metallicities. For each spaxel in the data cubes, we fit the continuum with a linear line between $0.75\mu m$ and $0.95\mu m$, excluding regions with strong emission lines. This is subtracted from the spectrum, allowing accurate fitting of the emission lines. We fit single Gaussian components to multiple emission lines (Hα, [NII]λλ6548,6584 and [SII]λλ6716,6731) simultaneously in each spaxel, using a least squares method. We assume that each of the emission lines trace regions with the same gas kinematics and hence have the same centroid and widths. Therefore, we fit the Hα lines, the strongest emission line in all of the galaxies, with three free parameters: centroid, width and line flux. For the remaining emission lines, the centroid and widths are fixed relative to the Hα line in the given spaxel, with the only free parameter being the emission line flux.

**Table 1.** Green pea targets, selected from Cardamone et al. (2009) to have bright (V<15 mag) and nearby (r<25″) stars that are observable from Palomar. Columns represent our ID, the SDSS ID of the galaxy, the RA and DEC of the galaxy, the redshift and r-band magnitude of the galaxy, the Hα flux of the galaxy within the SDSS fibre, SFR and stellar mass (where available from Cardamone et al. 2009).

| ID | SDSS | RA (deg) | Dec (deg) | z | mag (r-band) | Hα (erg/s/cm²) | SFR ($M_\odot yr^{-1}$) | Stellar Mass[a] ($M_\odot$) |
|---|---|---|---|---|---|---|---|---|
| GP1 | 587735695911747673 | 204.9196 | 55.46114 | 0.229 | 20.59 | $2.9\times10^{-15}$ | 3.22 (0.13)[a] | 9.94 |
| GP2 | 587741391573287017 | 145.9468 | 26.34516 | 0.237 | 19.98 | $1.3\times10^{-15}$ | 3.15 (0.07)[a] | 10.26 |
| GP3 | 588018253228081622 | 252.8746 | 25.71406 | 0.237 | 20.27 | $3.5\times10^{-15}$ | 4.72(0.10)[b] | |
| GP4 | 587739153352229578 | 117.9908 | 16.63701 | 0.265 | 20.25 | $4.0\times10^{-15}$ | 6.29 (0.21)[a] | 9.96 |
| GP5 | 587741534393598186 | 164.104 | 29.40817 | 0.19 | 20.40 | $4.3\times10^{-15}$ | 3.52 (0.05)[b] | |

[a] from Cardamone et al. (2009)

[b] derived from the Hα flux and the Kennicutt, Tamblyn & Congdon (1994) SFR relation.

**Table 2.** Properties of the SCAORS used for each green pea. Columns represent our ID, the RA and Dec of the SCAORS, the g-band magnitude of the SCAORS, and the separation between the galaxy and SCAORS on the sky.

| ID | RA$_\star$ (deg) | Dec$_\star$ (deg) | mag$_\star$ (g-band) | sep. (″) |
|---|---|---|---|---|
| GP1 | 204.9277 | 55.4646 | 11.560 | 20.6 |
| GP2 | 145.9388 | 26.3469 | 10.918 | 26.4 |
| GP3 | 252.8731 | 25.7210 | 14.138 | 25.3 |
| GP4 | 117.9881 | 16.6323 | 14.801 | 19.48 |
| GP5 | 164.0985 | 29.4039 | 14.800 | 23.36 |

We correct the velocity dispersion for instrumental dispersion by fitting a single Gaussian profile to a nearby sky line for each spaxel. We find that the average instrumental dispersion is $\sigma_{inst} \sim 50 km s^{-1}$. We combine this source of dispersion into the emission line fitting in order to extract the intrinsic Hα velocity dispersion in each spaxel.

We use published stellar masses from Cardamone et al. (2009). Since the photometry in the GPs is dominated by the strong nebular emission lines, standard spectral-energy-distribution (SED) fitting methods cannot be used with this data. Cardamone et al. (2009) instead use the SDSS spectra with the emission lines removed, convolved with 19 medium-band filters (Taniguchi 2004) to create synthetic medium-band photometric data. Combining this with near- and far- UV photometric data from GALEX (Martin et al. 2005) they then fit the SED with stellar population models (Maraston 1998, 2005) and correct for dust extinction using the

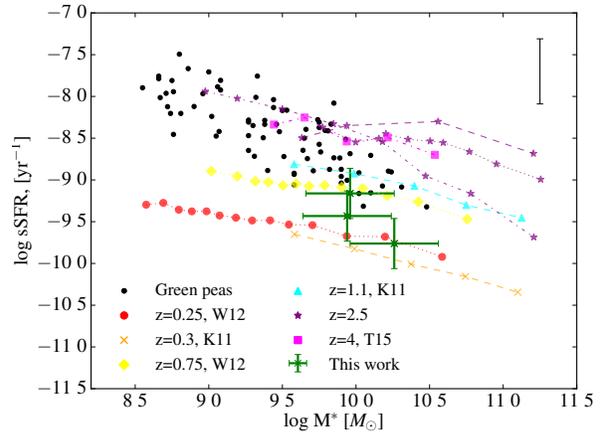

**Figure 2.** sSFR-mass sequence for green pea galaxies (black) with GP1, GP2 and GP4 indicated. GP5 is not shown as the stellar mass is not available from Cardamone et al. (2009). The population of green peas have some of the highest sSFRs at their redshift ($z \sim 0.2 - 0.3$). We also show the sSFR-Mass relation at a variety of redshifts from Tasca et al. (2015) (dash-dot line), Karim et al. (2011) (dashed line) and Whitaker et al. (2012) (dotted line). The typical spread in the sSFR-mass sequences is shown in the top right of the plot.

Calzetti et al. (2000) law, to obtain stellar mass estimates with typical errors of ∼0.3 dex.

## 3 RESULTS

We proceed by presenting the resolved spectroscopic properties of our four GPs. For all galaxies, we investigate their morphology from their normalized Hα counts and stellar continuum maps (Section 3.1), study their kinematics using maps of Hα velocities and velocity dispersions (Section 3.2) and the presence of AGN using emission line ratios (Section 3.4). For GP1 and GP2, which have higher S/N, we also investigate their metallicities (Section 3.3) and their electron densities (Section 3.5).

### 3.1 Normalized Hα counts

All of our GPs show high levels of Hα emission, as is expected, given their high SFRs and the fact that they have some of the highest sSFRs in the nearby Universe (Fig. 2). Fig. 4 shows normalized Hα maps for all four GPs. Both GP1 and GP2 show morphological asymmetry. In GP1 we see significant amounts of asymmetry towards the top of the galaxy, which is also seen in the SDSS image. In principle, this could result from the late stages of a merger, or it may be the result of patchy star formation, similar to that seen in high-redshift galaxies. To explore this further, we also study the stellar continuum map (Fig. 5), which traces the underlying structure of the stellar population, rather than just the areas with recent star formation. These are created by taking the median of the spectrum, with the emission lines masked, in each spaxel. We find that the stellar continuum also exhibits similar morphological disturbance, suggesting that the asymmetry seen in this galaxy is likely due to GP1 being the remnant of a recent merger.

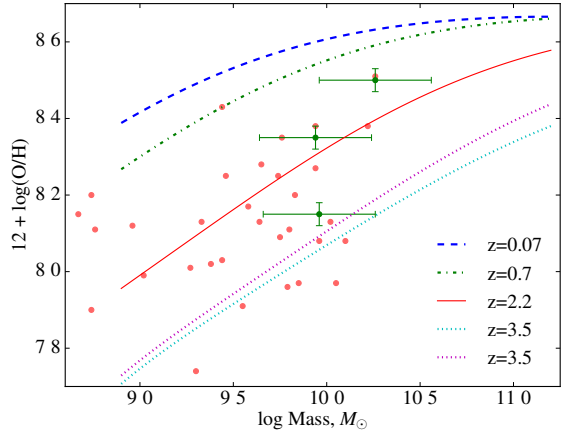

**Figure 3.** Metallicity vs stellar mass for GP1, GP2 and GP4 (green points with error bars) together with points from Amorín, Pérez-Montero & Vílchez (2010) (red). Also shown are the mass-metallicity relations, corrected to the PP04 metallicity calibration, at different redshifts from Maiolino et al. (2008). The GPs lie close to the higher redshift mass-metallicity relations at $z \sim 1-3$.

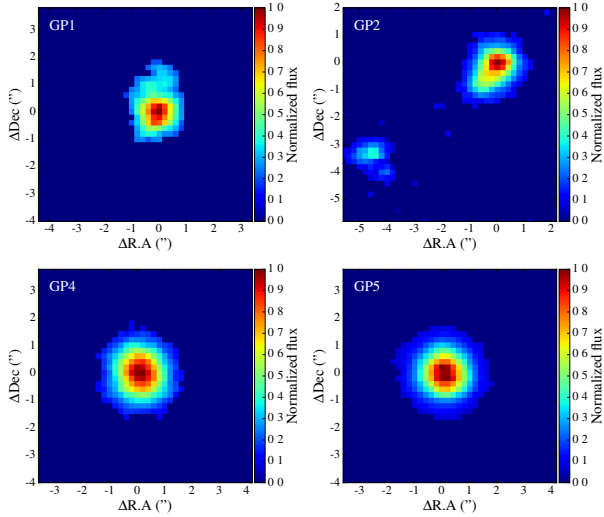

**Figure 4.** Normalized Hα counts for GP1, GP2, GP4 and GP5. GP1 and GP2 are both morphologically disturbed, indicating the presence of a recent merger (see text for more details). GP4 and GP5 show no signs of morphological disturbances. ΔR.A. and ΔDec are shown relative to the Hα peak in each GP.

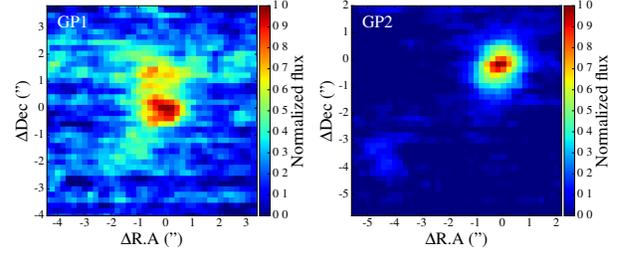

**Figure 5.** Normalized stellar continuum maps for GP1 and GP2. These are calculated by taking the median of each spectrum, with the emission lines masked. GP1 show significant morphological disturbances in the structure of the underlying stellar population, indicative of a recent merger whereas GP2 appears undisturbed. ΔR.A. and ΔDec are shown relative to the Hα peak in each GP.

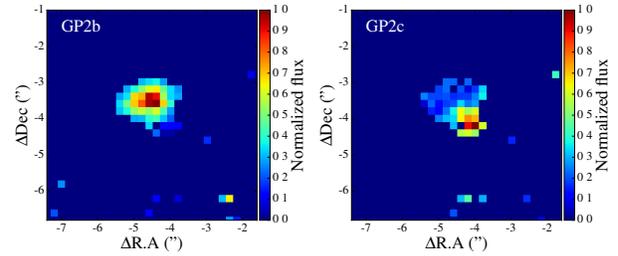

**Figure 6.** Normalized Hα counts for both components in GP2b/c, a secondary clump discovered in the image of GP2. ΔR.A. and ΔDec are shown relative to the Hα peak in GP2.

GP2 also shows some asymmetry in the normalized Hα map, towards the bottom left of the galaxy, similarly suggesting the presence of a recent merger or patchy star formation. However, in this case, neither the SDSS image (Fig 1) nor the stellar continuum map (Fig 5) show any morphological disturbance, indicating that this galaxy is unlikely to be a merger remnant. Alternatively, this elongation may be due to an ongoing interaction. In the data cube for GP2, we see a nearby clump towards the bottom left, ∼1 arcsecond in size, about 4 arcseconds (15 kpc) away from GP2 and at the same redshift as GP2. Fig. 6 shows the normalized Hα maps for this object. A single Gaussian is found not to be a good fit to the Hα emission line and instead a double Gaussian fit is used. This suggests that the clump is made of two components, hereafter referred to as GP2b and GP2c, at the same position but with separate kinematics. These components are potentially an ongoing merger itself (higher resolution IFS is required to explore this further). The elongation of GP2 observed in the Hα map may, therefore, be the result of star formation occurring in gas that resides in the tidal bridge between GP2 and its neighbour, rather than because this galaxy is a merger remnant. We find that the continuum flux ratio between GP2 and GP2b/c is $11 \pm 2$ to 1, consistent with this being a minor merger.

GP4 and GP5 have simple morphologies in the Hα emission, in agreement with their SDSS images (Fig. 1), with no evidence of any disturbances.

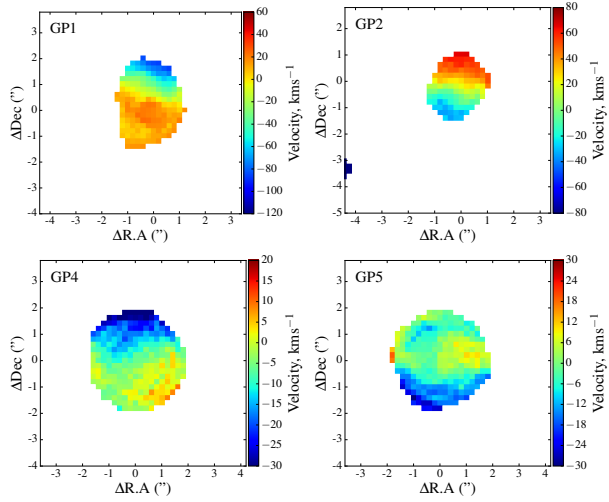

**Figure 7.** Velocity maps, calculated from the Hα line, for GP1, GP2, GP4 and GP5 and smoothed using a boxcar median method. Both GP1 and GP2 show the presence of a disc, while GP4 and GP5 do not show any coherent rotation. ΔR.A. and ΔDec are shown relative to the Hα peak in each GP.

## 3.2 Gas Kinematics

Fig. 7 shows the Hα velocity maps for each of our four GPs. In both GP1 and GP2, the Hα velocity maps show well-defined coherent rotation, i.e. the presence of a disk, with $V_{max}$ (the maximum rotation velocity seen in the outer regions of the galaxies) in the range 50-80 kms$^{-1}$ for both galaxies. Fig. 8 shows the velocities of each of the components in GP2b/c. There is some coherent rotation in these merging components, particularly in GP2c, which shows a velocity gradient. Unlike GP1 and GP2, the other two galaxies, GP4 and GP5, show only weak rotation, with maximum velocities up to 30 kms$^{-1}$. The velocity dispersion maps, measured from the Hα emission line and corrected for instrumental dispersion, are shown in Fig. 9 for all four of the GPs. Fig. 10 shows the Hα velocity dispersion for GP2b and GP2c. The velocity dispersion in all the GPs is $\sim 40 - 70$ kms$^{-1}$.

We compare the rotational velocities to the velocity dispersion by calculating $V_{max}/\sigma$ for each GP. $V_{max}$ is the maximum rotation velocity, seen in the outer regions of the galaxies, and $\sigma$ is the median velocity dispersion. We use the criterion, $V_{max}/\sigma > 1$ to identify galaxies that are rotationally supported and $V_{max}/\sigma < 1$ for dispersion-dominated systems. For GP1 and GP2, $V_{max}/\sigma$ is $\sim$1.8 and $\sim$3.4 respectively, i.e. these two galaxies are clearly rotationally supported. These values are similar to typical star-forming galaxies at high redshift (e.g the KROSS sample finds a median $V/\sigma = 2.2 \pm 1.4$, Stott et al. 2016) and consistent with other high-redshift galaxy populations studied by Swinbank et al. (2012, $z = 0.84 - 2.23$), Förster Schreiber et al. (2009, $z \sim 2$), Gnerucci et al. (2011, $z \sim 3$) and Epinat et al. (2012, $z > 0.8$). GP4 and GP5, on the other hand, are dispersion dominated, with $V_{max}/\sigma \sim 0.5$.

## 3.3 Metallicities

We investigate the metallicities in our sample of GPs using the [NII]λ6583/Hα emission-line ratio. This can be converted to the more commonly used $Z_{gas} = 12 + \log(O/H)$ metallicity using

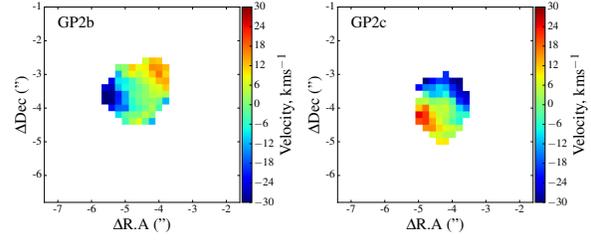

**Figure 8.** Velocity maps, calculated from the Hα line and smoothed using a boxcar median method, for GP2b and GP2c. ΔR.A. and ΔDec are shown relative to the Hα peak in GP2.

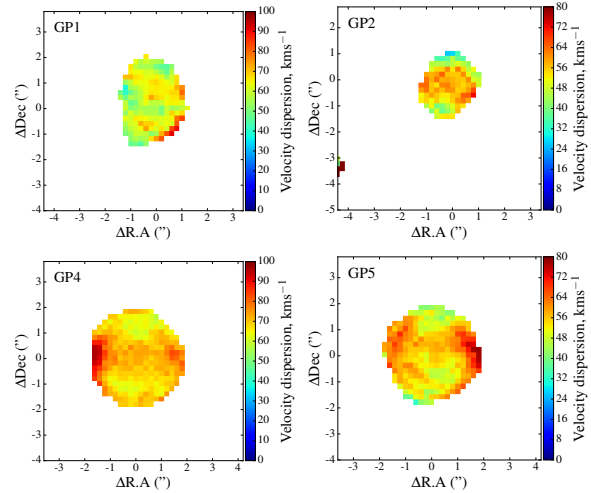

**Figure 9.** Velocity dispersion, calculated from the Hα line and smoothed using a boxcar median method, for GP1, GP2, GP4 and GP5. ΔR.A. and ΔDec are shown relative to the Hα peak in each GP.

the relation from Pettini & Pagel (2004) (PP04),

$$12 + \log(O/H) = 8.90 + 0.57 * \log([NII]\lambda6583/H\alpha)$$

Fig. 11 shows the metallicity maps for GP1, GP2 and GP4. All of these green peas have low metallicities of $12 + \log(O/H) \sim 8.35$ (25 per cent of solar) for GP1 and $\sim 8.5$ (40 per cent of solar) for GP2 and $12 + \log(O/H) \sim 8.15$ for GP4. The S/N of the [NII] emission line in GP5 is too low to create such a map, however we do measure the metallicity using an integrated spectrum and again find a sub-solar metallicity of $12 + \log(O/H) = 7.78 \pm 0.03$. The metallicities for all of these GPs are comparable to integrated metallicities found for other GPs in the literature. For example, Amorín, Pérez-Montero & Vílchez (2010) find that these galaxies have metallicities in the range $7.7 < Z_{gas} < 8.4$. In particular, they find that integrated metallicities of $12 + \log(O/H) = 8.38 \pm 0.05$ for GP1, $8.51 \pm 0.06$ for GP2 (via the method in Pérez-Montero & Contini (2009)), $7.88 \pm 0.25$ (via direct measurement) and $12 + \log(O/H) = 8.14 \pm 0.15$ (using the N2 index) for GP4, in agreement with the results found in this work. Note that large systematic discrepancies between metallicity estimates derived from different methods are seen in the literature. In particular, metallicities obtained via empirical calibrations and those based on theoretical methods can show discrepancies up to ∼0.6 dex (e.g. Kewley & Ellison 2008; Bian et al. 2017). In this work, we use the PP04 empirical calibration which is similar to the method used by Amorín, Pérez-Montero & Vílchez (2010) for GP1 and GP2, where we find very good agreement. Cardamone et al. (2009) find a mean metallicity of $Z_{gas} \sim 8.7$ for the GPs. However, the discrepancy between this result and the results found here (and in Amorín, Pérez-Montero & Vílchez (2010)) are not surprising as the Cardamone et al. (2009) results are obtained via a theoretical calibration (Kewley & Dopita 2002) using the [NII]λ6583/[OII]λλ3726,3729 ratio.

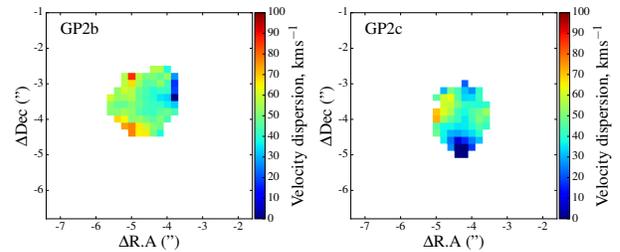

**Figure 10.** Hα Velocity dispersion for both GP2b and GP2c, smoothed using a boxcar median method ΔR.A. and ΔDec are shown relative to the Hα peak in GP2.

However, GP metallicities are consistently found to be sub-solar and, combined with their low masses and high SFRs, suggest that these galaxies are more similar to typical high-redshift galaxies than those in the local Universe. Fig. 3 shows the comparison of the metallicities and stellar masses calculated here to the mass-metallicity relations of Maiolino et al. (2008). These lines have been corrected to the PP04 metallicity calibration using the relations from Maiolino et al. (2008) and Pettini & Pagel (2004)). This indicates that the GPs in our study are similar to high-redshift galaxies at $z \sim 1 - 3$. Fig. 12 shows the radial profile of the gas phase metallicities for GP1 (black) and GP2 (red). Both GP1 and GP2 show broadly constant metallicities across the galaxy.

Fig. 13 shows the N/O ratios for GP1, GP2 and GP4 (bottom row). These values are calculated from the $\log([NII]\lambda\lambda6548,6584/[SII]\lambda\lambda6716,6731$ ratio (N2S2) using the empirical relation of Amorín, Pérez-Montero & Vílchez (2010),

$$\log(N/O) = -0.76 + 1.94N2S2 + 0.55(N2S2)^2$$

Due to the low S/N of the [NII] and [SII] in GP5, we calculate the log(N/O) ratio from the integrated spectra, and find a value of $-1.6 \pm 0.1$. For GP1, GP2 and GP4, the global values are $-1.20 \pm 0.06$, $-0.98 \pm 0.04$ and $-1.45 \pm 0.06$ respectively. All of the GPs have higher Nitrogen-to-Oxygen ratios(N/O) than typical star-forming galaxies at the same $12 + \log(O/H)$ metallicity. This is similar to the trends observed in previous studies of green peas (Amorín, Pérez-Montero & Vílchez 2010) and in high redshift galaxies relative to low redshift star forming galaxies(e.g. Masters et al. 2014; Shapley et al. 2015; Sanders et al. 2016). Comparing the global N/O values from Amorín, Pérez-Montero & Vílchez (2010) for GP1 and GP2 of $-1.3 \pm 0.1$ and $-1.1 \pm 0.1$ respectively, shows agreement with our results. For GP4, they find $\log(N/O) = -1.0 \pm 0.1$, which is not in agreement with our result of $-1.45 \pm 0.06$. Our map for GP4 shows radial variation in the log(N/O) values, with higher ratios in the central regions, up to ∼ $-1.3$, relative to the outskirts of the galaxy, although this is still much lower than that found in Amorín, Pérez-Montero & Vílchez (2010).

### 3.4 Presence of AGN

Past studies of GPs that have employed integrated spectra from the SDSS have found that these systems largely do not contain signs of AGN activity. For example, Cardamone et al. (2009) have shown that, out of the 103 GPs they studied, only 10 were AGN, 13 were classified as composite objects and 80 were star-forming, using standard emission-line diagnostics (e.g. Baldwin, Phillips & Terlevich 1981). We perform a similar analysis, but now in a resolved manner, to determine if any of GPs show AGN-like properties, particularly in the central regions, where an active AGN would be expected to dominate the ionization mechanisms. AGN enhance highly excited lines, e.g. [NII], [SII], relative to Hydrogen recombination lines such as Hα. Thus, for regions with AGN-like ionisation (Seyferts and LINERs), one expects to find values of $\log([NII]/H\alpha) > -0.22$ (e.g. Baldwin, Phillips & Terlevich 1981). We calculate this ratio for each spaxel in GP1, GP2 and GP4 to determine which regions of the galaxies may have AGN-like ionisation. These maps are shown in the left-hand column of Fig. 14. Although the S/N in GP5 is too low to perform this spatially resolved analysis, we are able to use the integrated spectra to calculate a global value of $\log([NII]/H\alpha) = -1.97 \pm 0.06$. We find that, in line with the findings of past work on GPs, none of our galaxies exhibit emission-line ratios that are typical of AGN-like ionization, even in the central regions where such line ratios would be expected if an AGN was present. Globally, the median $\log([NII]/H\alpha)$ values are very low compared to the AGN-SF division of -0.22, ranging between $-0.68 \pm 0.02$ (GP2) and $-1.97 \pm 0.06$ (GP5).

Similarly, the line ratio log[SII]/Hα can also be used as an indicator of the presence of an AGN, with values of $\log([SII]/H\alpha) > -0.3$ indicating the presence of an AGN (Kewley et al. 2006). Using this line ratio (right-hand plots in Fig. 14 and an integrated value for GP5 of $\log([SII]/H\alpha) = -1.36 \pm 0.03$), confirms the findings using $\log([NII]/H\alpha)$ - there is no evidence for an AGN in any of the GPs. These results indicate that AGN activity is not coeval with star formation in these galaxies.

### 3.5 Electron Density

We use the emission-line ratio $[SII]\lambda6716/[SII]\lambda6731$ to estimate the electron density in each spaxel for GP1 and GP2, via the relation from Osterbrock (1978):

$$[SII]\lambda6716/[SII]\lambda6731 = 1.50((1 + 1.38x)/(1 + 5.29x))$$

where $x = 10^{-2}N_e/T^{1/2}$, $N_e$ is the electron density and T is the electron temperature.

Using an electron temperature, for an ionized plasma, of ∼ $10^4 K$ (Kewley et al. 2015), we calculate the electron density in each spaxel for GP1 and GP2. For GP2, we see a broadly constant electron density across the galaxy of around 300 cm$^{-3}$. This is consistent with previous measurements of the electron densities in GPs (Amorín, Pérez-Montero & Vílchez 2010; Izotov, Guseva & Thuan 2011; Amorín et al. 2012a). In GP1, we see that the galaxy has much higher electron densities, of the order of 1000 cm$^{-3}$. However, these values are still consistent with the upper limits of what is expected for such systems (Izotov, Guseva & Thuan 2011).

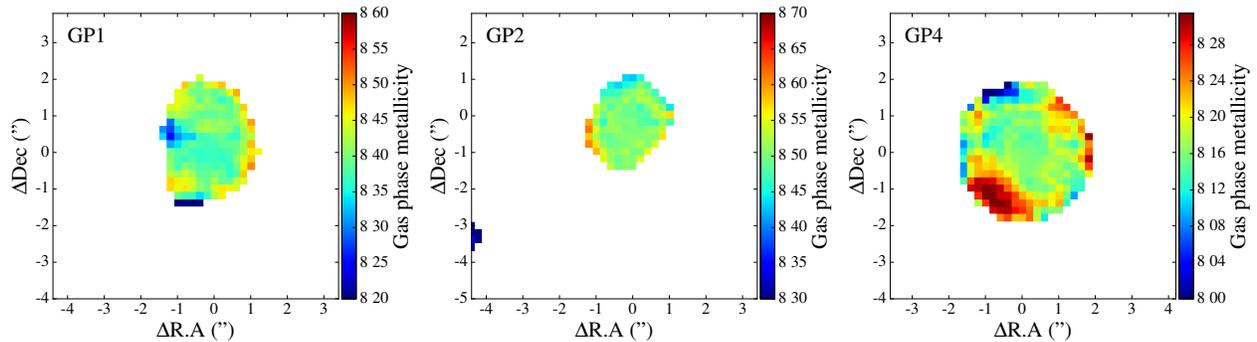

**Figure 11.** Gas phase metallicity, $12 + \log(O/H)$, maps of GP1, GP2 and GP4 derived from [NII]$\lambda$6583/H$\alpha$ emission line ratio and smoothed using a boxcar median method. $\Delta$R.A. and $\Delta$Dec are shown relative to the H$\alpha$ peak in each GP.

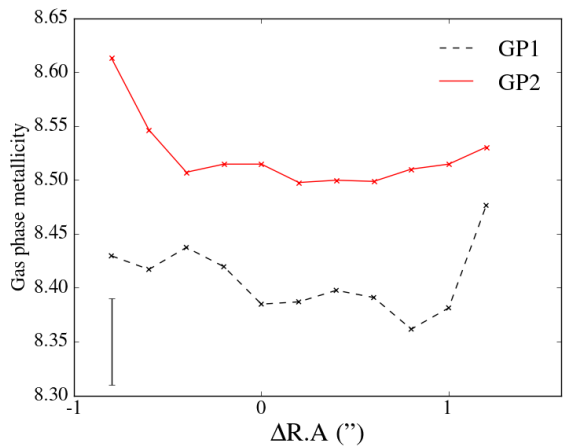

**Figure 12.** Gas phase metallicity, $12 + \log(O/H)$, along a horizontal line crossing the peak H$\alpha$ flux in each GP. The position, $\Delta$R.A. is shown relative to the maximum H$\alpha$ and a typical error is shown in the bottom right.

### 3.6 Outflows

Outflows are common in highly star-forming galaxies (e.g. Weiner et al. 2009; Bradshaw et al. 2013; Cicone, Maiolino & Marconi 2016). Amorín et al. (2012b) were the first to study the ionized gas kinematics of GPs and have studied outflows in GPs using long slit spectroscopy. They find that the galaxies can have complex emission-line profiles consisting of multiple kinematic components and, in at least two of these galaxies, the components are spatially resolved.

To investigate whether there is evidence for the presence of outflows in our GPs, we search for potential broad components in the H$\alpha$ emission. We repeat the emission-line fitting for the integrated spectra of each GP, adding an extra Gaussian component until the fit is no longer improved. Fig. 16 shows the improvement in the fits for the H$\alpha$ line in GP4 (this is typical of what we find for our other GPs). We find that the single Gaussian gives good agreement to the observed emission, although there is some excess emission in the broad wings of the emission line. While we don't find the same level of complexity as Amorín et al. (2012b), we find that by fitting an extra broad component gives a slightly better overall agreement to the observed emission line. This broad component has a FWHM of $\geqslant$200 km s$^{-1}$. We find similar results for all four of our GPs. They all show very good agreement with a single Gaussian, although an extra one or two Gaussian components can improve the fit slightly. While better data is needed to verify this, this provides some evidence that outflows may be present in our GPs, likely as the result of energetic feedback from supernovae due to their extreme levels of star formation.

## 4 DISCUSSION

In this section we collate the results presented above and investigate the nature of each of our GPs.

### 4.1 GP1

GP1 is relatively low mass ($10^{9.94}$ $M_\odot$; Cardamone et al. 2009), low metallicity ($12 + log(O/H) \sim 8.3 - 8.5$; Fig 11) and very highly star forming ($3.22 \pm 0.13$ $M_\odot yr^{-1}$; Cardamone et al. 2009). Its H$\alpha$ morphology (Fig. 4) shows asymmetry which could indicate that this galaxy is a recent merger remnant or that it may contain very patchy star formation. The SDSS image of this GP (Fig. 1) also shows morphological disturbances lending support to the merger-remnant hypothesis. Since we know that the photometry in at least one band is dominated by emission lines as a result of the high SFR, we also check the stellar continuum maps for this galaxy (Fig. 5). The stellar continuum map, which traces the underlying stellar structure rather than simply recent star formation, also shows significant morphological asymmetry. Therefore, it is likely that this system is the remnant of a recent interaction. However, the kinematics of this galaxy shows a coherently rotating disk with velocities up to 80 kms$^{-1}$ (Fig. 7). If a major merger caused the observed morphological disturbance in this galaxy, it would likely have destroyed any disk (Barnes & Hernquist 1991; Burkert & Naab 2003). Therefore, the presence of clear rotation, together with the morphological disturbance, suggests that any recent merger in this galaxy was most likely a minor merger (i.e. an interaction with a mass ratio $\lesssim$1:4).

In GP1, we measure low metallicities, with $Z_{gas} \sim 8.35$, approximately 25 per cent of the solar value. This is consistent with the metallicity measurements of Cardamone et al. (2009) who used the integrated flux in the SDSS fibre. Finally, we see no evidence in the [NII]/H$\alpha$ or [SII]/H$\alpha$ line ratios for any AGN-like ionization, indicating that this GP is purely star-forming and does not host any current AGN activity.

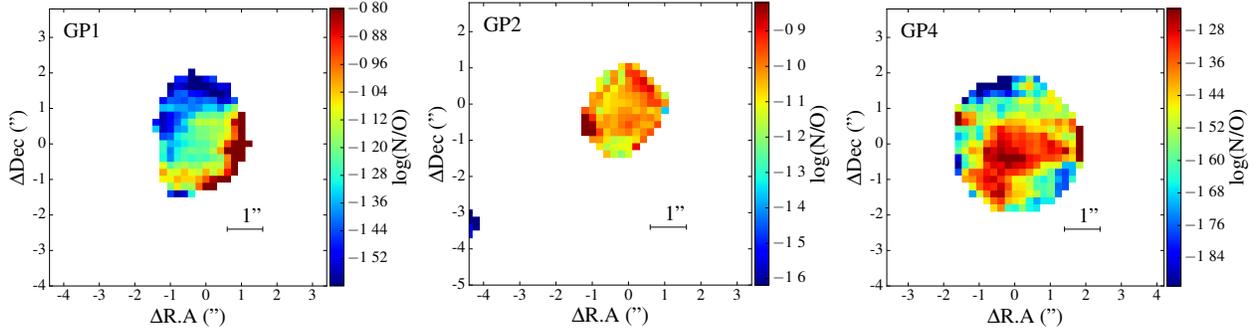

**Figure 13.** N/O ratio for GP1, GP2 and GP4 derived from the N2S2 parameter. ΔR.A. and ΔDec are shown relative to the Hα peak in each GP.

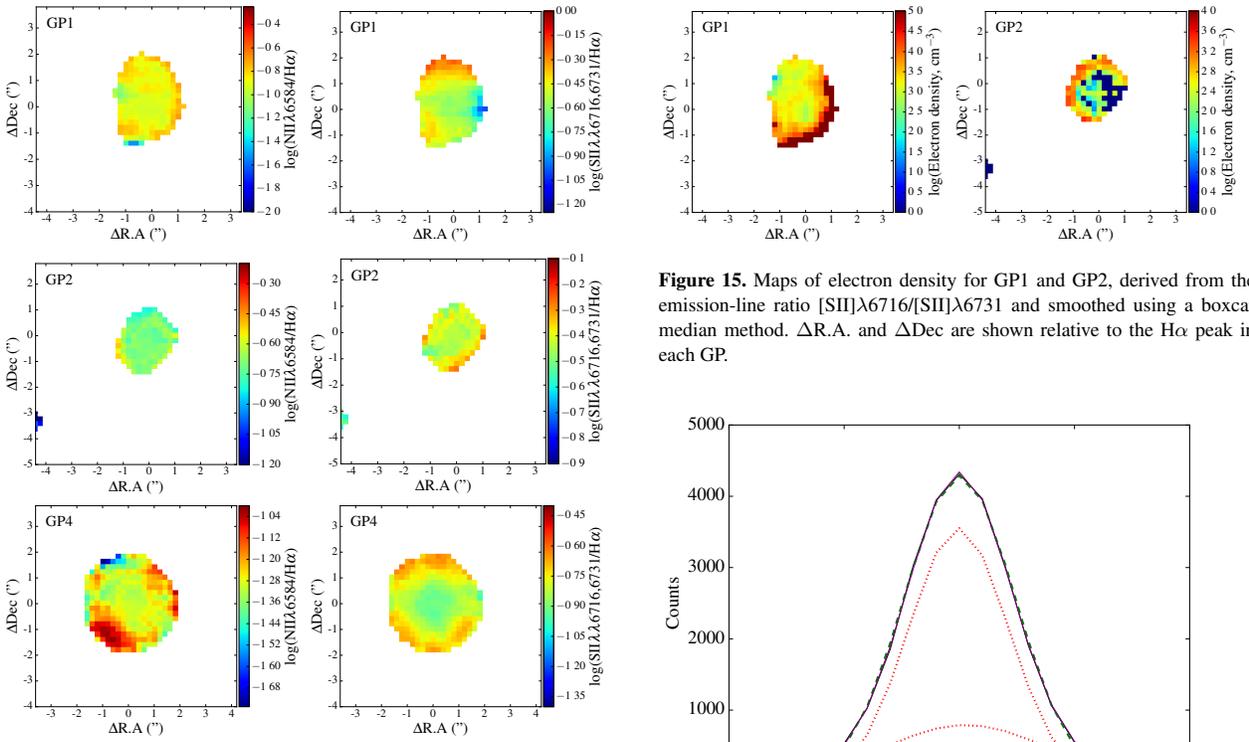

**Figure 14.** log([NII]/Hα) (left) and log([SII]/Hα) (right) maps for GP1, GP2 and GP4. $\log([NII]/H\alpha) > -0.22$ or $\log([SII]/H\alpha) > -0.3$ indicates the presence of AGN-like ionization. All spaxels show values below this level, indicating no AGN activity. ΔR.A. and ΔDec are shown relative to the Hα peak in each GP.

## 4.2 GP2

GP2 is a low metallicity ($12 + log(O/H) \sim 8.5$; Fig 11) galaxy, with a stellar mass of $10^{10.26}$ $M_\odot$ (Cardamone et al. 2009) and a high star formation rate ($3.15 \pm 0.07$ $M_\odot yr^{-1}$, Cardamone et al. 2009). While it does not show strong morphological disturbances either in its SDSS image or its stellar continuum map, it does show an elongation towards the lower left in the Hα map (Fig. 4). A small nearby neighbour is seen in the Hα map which is at the same redshift as GP2. Taken together, our analysis suggests that, while GP2 is not a merger remnant like GP1, it is likely to be involved in an ongoing minor merger with this smaller companion. Indeed, the elongation seen in the Hα map may be the result of star formation occurring in gas that resides in the tidal bridge between GP2 and its neighbour. Like in the case for GP1, this suggests that interactions may also be responsible for enhancing the star formation in this galaxy. The projected separation between GP2 and the neighbour, GP2b/c, is ∼15kpc. This is within the separations at which enhanced SFR is seen in close pairs. For example, Scudder et al. (2012) find average SFR enhancements of ~60 per cent for close pairs at r< $30h_{70}^{-1}$ kpc.

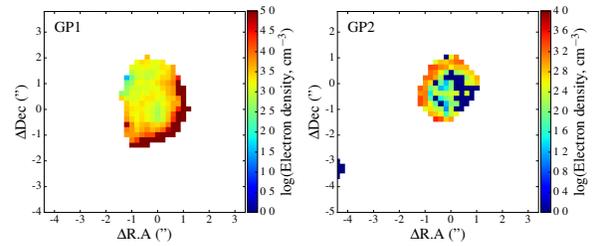

**Figure 15.** Maps of electron density for GP1 and GP2, derived from the emission-line ratio [SII]λ6716/[SII]λ6731 and smoothed using a boxcar median method. ΔR.A. and ΔDec are shown relative to the Hα peak in each GP.

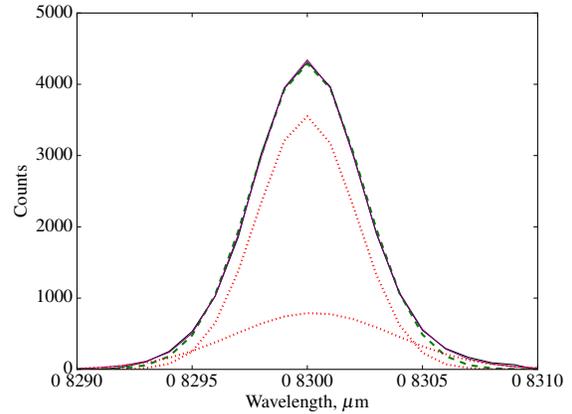

**Figure 16.** Fits to the observed Hα emission line in GP4 (solid black line). While a single Gaussian provides a very good fit to the observed emission lines (dashed green line), there is some excess emission in the broad wings of the emission line. Using multiple Gaussian components, each shown by the dotted red lines with the total shown by the solid purple line, provides a better fit to the observed emission line. This extra broad component suggests the presence of outflowing gas in this galaxy.

However, as in GP1, the presence of clear rotation (Fig. 7) in GP2 indicates that the interaction is not a major merger, in line with the size of the companion observed in the Hα maps. Together with the analysis above for GP1, these results suggest that major mergers are not a necessary condition for driving high SFRs in these types of galaxies, consistent with the findings of recent work (e.g. Dekel et al. 2009; Rodighiero et al. 2011; Conselice et al. 2013; Williams et al. 2014; Lofthouse et al. 2017; Fensch et al. 2017) and the lack of mergers seen at high-redshift (e.g. Stott et al. 2013). Finally, GP2 also shows no evidence of AGN-like ionization, indicating that the gas ionization is purely due to star formation activity.

### 4.3 GP4

GP4 is forming stars at a higher rate than GP1 or GP2 ($6.29 \pm 0.21 \, M_\odot yr^{-1}$; Cardamone et al. 2009), low metallicity ($12 + \log(O/H) \sim 8.15$; Fig. 11) and has a stellar mass of $10^{9.96} M_\odot$ (Cardamone et al. 2009). Unlike the previous two GPs it shows no morphological disturbances and appears to be a typical compact galaxy with no signs of any interactions (Fig 4). The high SFR in this undisturbed galaxy, compared to that in GP1 and GP2 which may partly be merger-driven, implies that mergers (including minor mergers) are not a necessary requirement for driving the extreme star formation properties in these types of galaxies and that secular processes are likely to be sufficient for driving strong star formation episodes. The galaxy does not show any clear rotation (Fig. 7) with rotational velocities less than ~ 30 kms$^{-1}$, and is dispersion dominated with V/σ ~ 0.5 (Fig. 9). This galaxy could still contain a disk but the rotation would need to be unresolved or the disk could be highly turbulent (Newman et al. 2013), which would produce the high velocity dispersion observed in the emission lines. Higher resolution IFU data, such as from HARMONI on the E-ELT (Thatte et al. 2014), is needed to investigate this further.

### 4.4 GP5

The final green pea, GP5, has a SFR of $3.52 \pm 0.05 M_\odot$ (calculated from the SDSS Hα flux using the SFR conversion from Kennicutt, Tamblyn & Congdon (1994)) and a low metallicity, calculated from the integrated spectra of $12 + \log(O/H) = 7.78 \pm 0.03$. It exhibits similar kinematic properties to GP4. The Hα morphology (Fig. 4) shows no signs of a recent merger, in agreement with the simple morphology seen in the SDSS images (Fig. 1). The Hα velocity map does not show any strong rotation (Fig. 7), with velocities less than ~ 30 kms$^{-1}$. It has V/σ ~ 0.5 indicating that this galaxy is dispersion dominated. As with the other three GPs, the [NII]/Hα or [SII]/Hα line ratios show no evidence of any AGN activity, indicating that this galaxy is currently purely star-forming.

## 5 SUMMARY

We have studied a sample of GPs which are likely to be local analogues of high-redshift star-forming galaxies. We have observed four GPs using the SWIFT+Palm3K IFU, to perform a spatially-resolved spectroscopic study of their kinematics and emission line properties. For each GP, we have fitted Hα, [NII] and [SII] emission lines in each spaxel, to derive maps of the normalized Hα flux, velocity, velocity dispersion, metallicity, gas-ionization mechanism and electron density. We have used these maps to explore the physical properties of these galaxies, study the processes that are influencing their stellar mass growth and the potential role of AGN in their evolution. Our main results can be summarized as follows:

● Two of our GPs show morphologies indicative of recent or ongoing mergers. GP1 appears to be the remnant of a recent merger, while GP2 has a small, nearby interacting neighbour. This suggests that interactions may be responsible for the high SFRs observed in these galaxies. However, GP4 and GP5 show no signs of a recent interaction (and have similar SFRs to GP1 and GP2), indicating that a merger is not a *necessary* requirement for driving the high star formation in these types of galaxies.

● By creating maps of the Hα velocity and velocity dispersion, we conclude that GP1 and GP2 are rotationally supported, while GP4 and GP5 are dispersion dominated. Interestingly, the rotationally supported galaxies are the ones which show morphological disturbances. The fact that there is clear rotation in these galaxies, even when they are morphologically disturbed, implies that any recent mergers or interactions are likely to have low mass ratios (i.e. the interactions are minor mergers). It is possible, therefore, that the high SFRs observed in such systems could partly be the result of SFR-enhancement by minor mergers, as is also observed in systems in the nearby Universe (e.g. Kaviraj 2014b,a).

● Using the $[NII]\lambda6583/H\alpha$ emission line ratio to create resolved maps of the 12+log(O/H) metallicity, we confirm that these galaxies have low metallicities, $Z_{gas} \sim 8.3 - 8.5$. (25 to 40 per cent solar).

● We find no evidence for ongoing AGN activity in any of the GPs in either the $\log([NII]/H\alpha)$ or the $\log([SII]/H\alpha)$ ratios, indicating that AGN activity is not coeval with star formation in these systems.

● For GP1 and GP2 we calculate the resolved electron densities. We find that they both have broadly constant electron densities, with GP2 showing values of ~ 300 cm$^{-3}$ and GP1 showing high values of the order of 1000 cm$^{-3}$ (but which are, nevertheless, consistent with the expectations for galaxies with similar masses). As forthcoming and future instruments (e.g. MaNGA, WEAVE, HARMONI) open up the nearby and distant Universe to IFU studies, similar analysis to that presented in this paper will enable us to use such instruments to explore the role of mergers in driving stellar mass growth and the role of AGN activity in regulating the evolution of the observable Universe over cosmic time.

## ACKNOWLEDGEMENTS

EKL acknowledges support from the UK's Science and Technology Facilities Council [grant number St/K502029/1]. SK acknowledges a Senior Research Fellowship from Worcester College Oxford. RCWH was supported by the Science and Technology Facilities Council [STFC grant numbers ST/H002456/1, ST/K00106X/1 & ST/J002216/1]. We are grateful to John Stott for many constructive comments on this analysis and to Ricardo Amorín for many helpful comments and providing measurements for comparison in this work.

## REFERENCES

Amorín R. et al., 2017, Nature Astronomy, 1, 0052

Amorín R., Pérez-Montero E., Vílchez J. M., Papaderos P., 2012a, ApJ, 749, 185

Amorín R., Vílchez J. M., Hägele G. F., Firpo V., Pérez-Montero E., Papaderos P., 2012b, ApJL, 754, L22

Amorín R. O., Pérez-Montero E., Vílchez J. M., 2010, ApJL, 715, L128

Baldwin J. A., Phillips M. M., Terlevich R., 1981, PASP, 93, 5

Barnes J. E., Hernquist L. E., 1991, ApJL, 370, L65

Bian F., Kewley L. J., Dopita M. A., Blanc G. A., 2017, ApJ, 834, 51

Bian F., Kewley L. J., Dopita M. A., Juneau S., 2016, ApJ, 822, 62

Bradshaw E. J. et al., 2013, MNRAS, 433, 194

Burkert A., Naab T., 2003, in Lecture Notes in Physics, Berlin Springer Verlag, Vol. 626, Galaxies and Chaos, Contopoulos G., Voglis N., eds., pp. 327–339

Calzetti D., Armus L., Bohlin R. C., Kinney A. L., Koornneef J., Storchi-Bergmann T., 2000, ApJ, 533, 682

Cardamone C. et al., 2009, MNRAS, 399, 1191

Cicone C., Maiolino R., Marconi A., 2016, A&A, 588, A41

Conselice C. J., Mortlock A., Bluck A. F. L., Grützbauch R., Duncan K., 2013, MNRAS, 430, 1051

Cowie L. L., Barger A. J., Songaila A., 2016, ApJ, 817, 57

Daddi E. et al., 2010, ApJL, 714, L118

De Barros S., Reddy N., Shivaei I., 2016, ApJ, 820, 96

Dekany R. et al., 2013, ApJ, 776, 130

Dekel A. et al., 2009, Nature, 457, 451

Epinat B. et al., 2012, A&A, 539, A92

Faisst A. L. et al., 2016, ApJ, 821, 122

Fensch J. et al., 2017, MNRAS, 465, 1934

Förster Schreiber N. M. et al., 2009, ApJ, 706, 1364

Gnerucci A. et al., 2011, A&A, 528, A88

Greis S. M. L., Stanway E. R., Davies L. J. M., Levan A. J., 2016, MNRAS, 459, 2591

Hopkins A. M., Beacom J. F., 2006, ApJ, 651, 142

Izotov Y. I., Guseva N. G., Thuan T. X., 2011, ApJ, 728, 161

Karim A. et al., 2011, ApJ, 730, 61

Kaviraj S., 2014a, MNRAS, 440, 2944

Kaviraj S., 2014b, MNRAS, 437, L41

Kaviraj S. et al., 2013, MNRAS, 429, L40

Kennicutt, Jr. R. C., Tamblyn P., Congdon C. E., 1994, ApJ, 435, 22

Kewley L. J., Dopita M. A., 2002, ApJS, 142, 35

Kewley L. J., Ellison S. L., 2008, ApJ, 681, 1183

Kewley L. J., Groves B., Kauffmann G., Heckman T., 2006, MNRAS, 372, 961

Kewley L. J., Zahid H. J., Geller M. J., Dopita M. A., Hwang H. S., Fabricant D., 2015, ApJL, 812, L20

Lofthouse E. K., Kaviraj S., Conselice C. J., Mortlock A., Hartley W., 2017, MNRAS, 465, 2895

Madau P., Pozzetti L., Dickinson M., 1998, ApJ, 498, 106

Maiolino R. et al., 2008, A&A, 488, 463

Maraston C., 1998, MNRAS, 300, 872

Maraston C., 2005, MNRAS, 362, 799

Martin D. C. et al., 2005, ApJL, 619, L1

Maseda M. V. et al., 2014, ApJ, 791, 17

Masters D. et al., 2014, ApJ, 785, 153

Newman S. F. et al., 2013, ApJ, 767, 104

Osterbrock D. E., 1978, Phys. Scr., 17, 285

Overzier R., Lemson G., Angulo R. E., Bertin E., Blaizot J., Henriques B. M. B., Marleau G.-D., White S. D. M., 2013, MNRAS, 428, 778

Pérez-Montero E., Contini T., 2009, MNRAS, 398, 949

Pettini M., Pagel B. E. J., 2004, MNRAS, 348, L59

Rodighiero G. et al., 2011, ApJL, 739, L40

Sanders R. L. et al., 2016, ApJ, 816, 23

Scudder J. M., Ellison S. L., Torrey P., Patton D. R., Mendel J. T., 2012, MNRAS, 426, 549

Shapley A. E. et al., 2015, ApJ, 801, 88

Stott J. P., Sobral D., Smail I., Bower R., Best P. N., Geach J. E., 2013, MNRAS, 430, 1158

Stott J. P. et al., 2016, MNRAS, 457, 1888

Swinbank A. M., Sobral D., Smail I., Geach J. E., Best P. N., McCarthy I. G., Crain R. A., Theuns T., 2012, MNRAS, 426, 935

Taniguchi Y., 2004, in Studies of Galaxies in the Young Universe with New Generation Telescope, Arimoto N., Duschl W. J., eds., pp. 107–111

Tasca L. A. M. et al., 2015, A&A, 581, A54

Thatte N., Tecza M., Clarke F., Goodsall T., Lynn J., Freeman D., Davies R. L., 2006, in Proc. SPIE, Vol. 6269, Society of Photo-Optical Instrumentation Engineers (SPIE) Conference Series, p. 62693L

Thatte N. A. et al., 2014, in Proc. SPIE, Vol. 9147, Ground-based and Airborne Instrumentation for Astronomy V, p. 914725

van der Wel A. et al., 2011, ApJ, 742, 111

van Dokkum P. G., 2001, PASP, 113, 1420

Weiner B. J. et al., 2009, ApJ, 692, 187

Whitaker K. E., van Dokkum P. G., Brammer G., Franx M., 2012, ApJL, 754, L29

Williams C. C. et al., 2014, ApJ, 780, 1

This paper has been typeset from a TEX/ LATEX file prepared by the author.